\documentclass[epjCONF]{svjour}
\usepackage{amsmath,amssymb}
\usepackage{graphicx}
\usepackage[varg]{txfonts} 
\usepackage{mathrsfs}
\newcommand{\mrs}{\mathscr}

\usepackage[latin1]{inputenc}
\session-title{%
19$^{\textnormal{\footnotesize th}}$ International %
IUPAP Conference on Few-Body Problems in Physics%
}
\begin{document}
\title{\boldmath$NN$ Interaction JISP16: Current Status and Prospect}
\author{A. M. Shirokov\inst{1,2}\fnmsep\thanks{\email{shirokov@nucl-th.sinp.msu.ru}}
\and V. A. Kulikov\inst{1}
\and P. Maris\inst{2}
\and A. I. Mazur\inst{3}
\and E. A. Mazur\inst{3}
\and J. P. Vary\inst{2}}
\institute{Skobeltsyn Institute of Nuclear Physics, Moscow State University,
 Moscow 119991 Russia
\and Department of Physics and Astronomy, Iowa State University, Ames,
IA 50011, USA
\and Pacific National University, 136 Tikhookeanskaya, Khabarovsk 680035, Russia} 
\abstract{We discuss realistic nonlocal $NN$ interactions of a new
type --- $J$-matrix Inverse Scattering Potential (JISP). In an {\em ab
  exitu} approach, these 
interactions are fitted to not only two-nucleon data ($NN$
scattering data and deuteron properties) but also to the properties of
light nuclei without referring to three-nucleon forces. We
discuss recent progress with the  
{\em ab initio} No-core Shell Model (NCSM) approach 
and respective progress in developing {\em ab exitu}
JISP-type $NN$-interactions
together with plans of their forthcoming improvements.} 

\maketitle

Significant progress was achieved\ \ within\ \ the last decade in {\em ab
initio} studies of light nuclei. Nowadays, due to increased computing
power and novel techniques, {\em ab initio} approaches like the No-core
Shell Model (NCSM) \cite{NCSM}, the Green's function Monte Carlo
\cite{GFMC-rev} and the coupled-cluster theory  \cite{CCDean} are able
to reproduce properties of a large number of atomic nuclei with mass up
to $A=16$ and can be extended for heavier nuclei.

The {\em ab initio} methods require a
reliable realistic strong interaction providing an accurate
description of $NN$ scattering data and high-quality predictions
for binding energies, spectra and other observables in light nuclei. A
number of meson-exchange  potentials sometimes supplemented with
phenomenological terms to  achieve high accuracy in fitting $NN$ data
(CD-Bonn \cite{Bonn},   Nijmegen \cite{Stocks}, Argonne \cite{Argonne})
have been developed that should be used together with modern $NNN$ forces 
(Urbana \cite{Urbana,GFMC}, {Illinois} \cite{Illinois},
Tucson--Melbourne \cite{TM,TM-pr}) to reproduce properties of many-body 
nuclear systems. On the other hand, one sees the emergence of \ $NN$ \ and \
$NNN$ \ interactions with ties to QCD \cite{Bedaque,Bochum,Idaho,NVGON}. 

Three-nucleon forces require a significant increase of computational
resources needed to diagonalize a many-body Hamiltonian matrix since the
$NNN$ interaction increases   the
number of non-zero matrix elements approximately  by a factor of 30 in
the case of $p$-shell nuclei. As a result, one needs to restrict the
basis space in many-body calculations 
when $NNN$ forces are involved that makes the predictions less
reliable. {\em Ab initio} many-body studies benefit from the use of
recently developed purely two-nucleon interactions of INOY  (Inside
Nonlocal Outside Yukawa) \cite{Doleschall,Plessas} and JISP ($J$-matrix
Inverse Scattering Potential) \cite{ISTP,JISP6,JISP16,JISP16_web} 
types fitted not only to the $NN$ data but also to binding energies of
$A=3$ and heavier nuclei. At the fundamental level, these $NN$
interactions are supported by the work  of 
Polyzou and Gl\"ockle who demonstrated \cite{Poly} that
a realistic $NN$ interaction is equivalent at the $A=3$ level to some
$NN + NNN$ interaction where the new $NN$ force is
related to the initial one through a phase-equivalent transformation (PET).
It seems reasonable then to exploit this freedom and work 
to minimize the need for the explicit introduction 
of three and higher body forces. Endeavors along these lines have
resulted in the design of 
INOY and JISP strong interaction models.

We discuss here the progress in development of the JISP $NN$
interactions and related progress in NCSM studies of light nuclei.

\section{The original JISP16}

The $J$-matrix inverse scattering approach was suggested in
Ref. \cite{Ztmf1}. It was further developed and used to design a
high-quality JISP $NN$ interaction in Ref. \cite{ISTP}. A nonlocal interaction
obtained in this approach is in the form of a matrix in the oscillator
basis in each of the $NN$ partial waves. To reproduce scattering data in
a wider energy range, one needs to 
increase the size of the potential matrix and/or the $\hbar\Omega$ parameter
of oscillator basis. From the point of view of shell model applications,
it is desirable however to reduce the size of potential matrices and to
use $\hbar\Omega$ values in the range of few tens of MeV. A compromise
solution is to use $\hbar\Omega=40$~MeV with $N_{\max} = 9$ truncation
of potential matrices \cite{ISTP}, i.e., we use   potential matrices of the rank
$r=5$ in $s$ and $p$ $NN$ partial waves, $r=4$ matrices in $d$ and $f$
partial waves, etc.; in the case of coupled waves, the rank of the
potential matrix is a sum of of the respective  ranks, e.g., the rank of
the coupled $sd$ wave matrix is $r=5+4=9$. The $N_{\max}=9$ truncated
JISP interaction with $\hbar\Omega=40$~MeV  provides an excellent description
of $NN$ scattering data with $\rm \chi^2/datum = 1.03$ for
the 1992 $np$ data base (2514 data), and 1.05 
for the 1999 $np$ data base (3058 data) \cite{chi2-priv}.

\begin{table*}
\caption{JISP6 and JISP16 deuteron property predictions  in
comparison with the ones obtained with various realistic potentials.}
\label{d-prop}
\begin{center}
\begin{tabular}{ccccccc} \hline
Potential & $E_{d}$, MeV &\parbox{2.1cm}{\centering$d$ state probability,~\%}
& \parbox{1.5cm}{\centering rms radius,~fm} & $Q$, fm$^2$
                   & \parbox{2.1cm}{\centering As.$^{\vphantom{l}}$ norm. const.
                    ${\mrs A}_s^{\vphantom{l}}$, fm$^{-1/2}_{\vphantom{q}}$}
&$\displaystyle\eta=\frac{{\mrs A}\vphantom{'}_d}
           {{\mrs A}_{s\vphantom{_a}}}$\\ \hline
JISP6, JISP16 & $-2.224575$ & 4.1360 & 1.9647 &0.2915 &0.8629 &0.0252\\
  Nijmegen-II & $-2.224575$ & 5.635 & 1.968 & 0.2707 &0.8845 & 0.0252\\
AV18 & $-2.224575$ &5.76 & 1.967 & 0.270 & 0.8850 & 0.0250\\
CD--Bonn & $-2.224575$ & 4.85 & 1.966 & 0.270 & 0.8846 &0.0256\\
Nature 
& $-2.224575(9)$ & --- 
&1.971(6)
&0.2859(3)  &0.8846(9) &0.0256(4)         
\\ \hline
\end{tabular}
\end{center}
\end{table*}

PETs originating from unitary transformations of the oscillator basis
proposed in \cite{PHT,Halo-Ann},
give rise to ambiguities of interaction obtained in the $J$-matrix
inverse scattering approach. These ambiguities are eliminated at the first
stage by postulating the simplest tridiagonal form of the $NN$
interaction in uncoupled and quasi-tridiagonal form in coupled $NN$
partial waves \cite{ISTP}. At the next stage, PETs are used to fit the
JISP interaction to various nuclear properties. First of all, the $sd$
component of the $NN$ interaction is modified with the
help of PETs to reproduce the deuteron quadrupole moment $Q$ and rms radius
without violating the excellent description of scattering data. The
deuteron property predictions obtained with JISP and other interaction
models are listed in Table \ref{d-prop}. It is worth noting here that
the deuteron binding energy $E_d$ and asymptotic normalization constants
${\mrs A}_s$ and ${\mrs A}_d$ are used as an input in the inverse
scattering approach and are not affected by PETs.

After that we employ PETs in other $NN$ partial waves attempting to improve
the description of binding energies and spectra of light nuclei in
NCSM calculations.  Following this {\em ab exitu} route, the JISP6 $NN$ interaction
fitted to properties of nuclei with masses $A\leq 6$, was proposed
\cite{JISP6}. It was found out later that JISP6 strongly overbinds
nuclei with $A\geq 10$.  Therefore a new fit of PET parameters was
performed that resulted in the JISP16 interaction \cite{JISP16} fitted
to nuclei with masses up through $A=16$.  After discussing methods
used in the fit of the JISP16 interaction and results obtained with
it, we shall concentrate on some drawbacks of this interaction
revealed due to the recent progress in the NCSM approach \cite{NCFC}
and attainment of larger basis spaces in calculations, and propose its new
refined version JISP16$_{2010}$.

%
%

\begin{table}
\tabcolsep=2pt
\caption{Binding energies (in MeV) of nuclei obtained with LSO(2)
renormalized JISP16 (``effective interaction'', from
Ref.~\cite{JISP16}), and variational lowerbounds from more recent
calculations, generally obtained in larger model spaces than was
feasible in 2006, when JISP16 was developed.
\label{bind16}}
\begin{center}
\begin{tabular}{cccccc}\hline
  Nucleus & Nature & LSO renormalized    & $N_{\max}$ & Variational & $N_{\max}$  \\
          &        & from Ref.~\cite{JISP16}&         & lowerbound  &          \\ \hline
  $^3$H   &  8.482 &  8.496 &  14  &  8.367 & 20 \\
  $^3$He  &  7.718 &  7.797 &  14  &  7.663 & 20 \\
  $^4$He  & 28.296 & 28.374 &  14  & 28.299 & 18 \\
  $^6$He  & 29.269 & 28.32  &  12  & 28.616 & 16 \\
  $^6$Li  & 31.995 & 31.00  &  12  & 31.340 & 16 \\
  $^7$Li  & 39.245 & 37.59  &  10  & 37.954 & 12 \\
  $^7$Be  & 37.600 & 35.91  &  10  & 36.273 & 12 \\
  $^8$Be  & 56.500 & 53.40  &   8  & 53.731 & 10 \\
  $^9$Be  & 58.165 & 54.63  &   8  & 53.577 &  8 \\
  $^9$B   & 56.314 & 52.53  &   8  &   51.308  & 8  \\
$^{10}$Be & 64.977 & 61.39  &   8  & 60.596 &  8 \\
$^{10}$B  & 64.751 & 60.95  &   8  & 60.455 &  8 \\
$^{10}$C  & 60.321 & 56.36  &   8  &   55.264  & 8   \\
$^{11}$B  & 76.205 & 73.0   &   6  & 69.182 & 6   \\
$^{11}$C  & 73.440 & 70.1   &   6  & 66.060 &6   \\
$^{12}$B  & 79.575 & 75.9   &   6  &  71.190 &6  \\
$^{12}$C  & 92.162 & 91.0   &   6  & 92.814 & 10 \\
$^{12}$N  & 74.041 & 70.2   &   6  &  64.539 &6  \\
$^{13}$B  & 84.453 & 82.1   &   6  &  73.527 & 6  \\
$^{13}$C  & 97.108 & 96.4   &   6  & 93.208 & 6   \\
$^{13}$N  & 94.105 & 93.1   &   6  &   89.690 & 6 \\
$^{13}$O  & 75.558 & 72.9   &   6  & 69.066 &  8 \\
$^{14}$C  &105.285 &106.0   &   6  &106.853 &  8 \\
$^{14}$N  &104.659 &106.8   &   6  &109.136 &  8 \\
$^{14}$O  & 98.733 & 99.1   &   6  & 99.337 &  8 \\
$^{15}$N  &115.492 &119.5   &   6  & 114.409 & 6   \\
$^{15}$O  &111.956 &115.8   &   6  & 110.139 & 6   \\
$^{16}$O  &127.619 &133.8   &   6  &134.494 &  8 \\ \hline
\end{tabular}
\end{center}
\end{table}

Our fitting procedure was one of `trial-and-error' where we worked
with only a few partial waves that we thought might be important for
light nuclei.  We fitted only the excitation energies of the lowest
$^6$Li levels and the $^6$Li and $^{16}$O binding energies.  To save
time, we performed the NCSM calculations in small enough
$N_{\max}\hbar\Omega$ basis spaces
(up to $N_{\max}=10$ for $^6$Li and up to $N_{\max}=4$ for $^{16}$O)
using the effective interaction obtained from the original `bare' inverse
scattering potential by means of
the Lee--Suzuki--Oka\-mo\-to (LSO) renormalization
procedure \cite{LSO}.  The renormalization procedure was truncated at
the two-body cluster level --- i.~e. induced three-body, four-body,
etc., contributions are neglected; hence we refer to these
calculations as LSO(2) renormalized.  The variational principle holds
for the bare interaction results, but not for results obtained with
the LSO(2) renormalized interaction.  Conventional wisdom suggests
that the minimum with respect to $\hbar\Omega$ of the ground state energies
obtained with the LSO(2) renormalized interaction provides the best
estimate of the actual ground state energy; furthermore, 
this minimum increases with increasing $N_{\max}$ in a number of
nuclei, at least for 
relatively small 
basis spaces.  It was therefore believed that this
minimum provides a reasonable lower bound for the actual ground state
energy.  On the other hand the minimum obtained with the bare
interaction provides us with a strict upper bound, because of the
variational principle.  After obtaining a reasonable description of
the lowest $^6$Li levels and the $^6$Li and $^{16}$O binding energies,
we checked that the binding energies and spectra of all the remaining $s$
and $p$ shell nuclei are well-described in similarly small model
spaces.

The results presented in Table~\ref{bind16} are obtained in the {\it
ab initio} NCSM calculations with the obtained $NN$ interaction, the
{\it ab exitu} JISP16, in larger model spaces.  This description of
the binding energies is somewhat worse than the one obtained during
the fit in smaller model spaces, but is still very reasonable.
However, moving to larger model spaces revealed that the convergence
of the LSO(2) renormalized JISP16 interaction is not as uniform as
suggested by conventional wisdom.  In particular for a number of heavier
nuclei with $A \ge 12$, the current variational upper bound tends to be
lower than our best estimates 3 years ago.

\begin{figure}
\centerline{\includegraphics[width=\columnwidth]{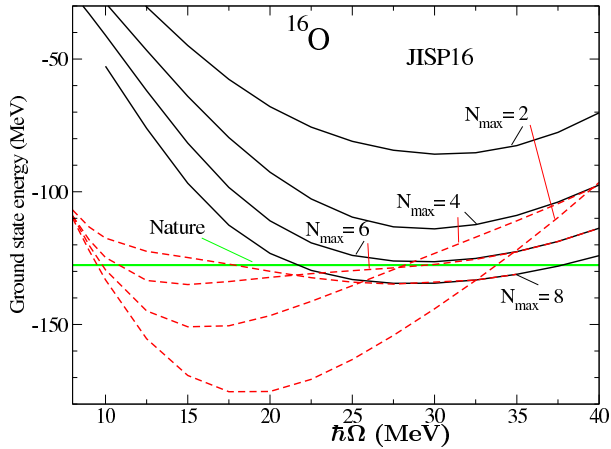}}
\caption{Results for the ground state energy of $^{16}$O with bare
    (solid) and LSO(2) renormalized (dashed) JISP16 interaction as a
    function of the oscillator parameter $\hbar\Omega$. When JISP16
    was developed \cite{JISP16}, the largest model space calculations
    where $N_{\max}$ = 6. \label{hwdep}}
\end{figure}

We illustrate this issue by the $^{16}$O ground state energy
calculations in Fig. \ref{hwdep}.  The solid curves are our NCSM
results with the bare interaction JISP16.  The variational principle
holds for the bare interaction results; hence these 
curves form
a strict upper bound for the ground state energy.  The dashed curves in
Fig.~\ref{hwdep} were obtained in the more conventional NCSM
calculations with the LSO(2) renormalized interaction derived from the
initial bare interaction JISP16.  For the series of calculations with
$N_{\max}=0$  (not shown), 2, 4, and 6, the minimum of these dashed
curves increases with model spaces, and at $N_{\max}=6$ this minimum
is only slightly below the corresponding minimum obtained with the
bare interaction ($-133.8$ MeV and $-126.2$~MeV respectively). However,
the $N_{\max}=8$ curves for \ the \ LSO(2) \ and \ bare \ interaction \
(dashed \ and \ solid
curves) clear\-ly demonstrate that with the LSO(2)
renormalized interaction does not converge uniformly to the infinite
basis space results, nor does it necessarily produce a lower bound.

Similar trends were found for most of the $p$ shell nuclei: the LSO(2)
renormalized (or effective) interactions produce results which are
neither an upper bound nor a lower bound, and the approach to the
infinite basis space is non-monotonic.  Hence the convergence pattern
of the effective interaction results is difficult to assess.
Nevertheless, in small model spaces they do give a reasonable estimate
of the actual binding energies.  As such, the LSO procedure is very
useful, in particular for interactions that converge only at very
large model spaces, such as N3LO, CD-Bonn, or Argonne.
However, JISP16 is a very soft interaction, and the results with the
bare interaction converge rapidly, making it unnecessary to use the
LSO renormalization techniques for nuclei up to $A \sim 16$.  Indeed,
this is confirmed by the fact that for large values of $\hbar\Omega$,
the bare and LSO(2) renormalized interaction lead to identical
results, and that the  $\hbar\Omega$ range  over which the results
from the bare and LSO(2) renormalized interactions coincide
increases 
with the basis space.

The nuclear Hamiltonian based on the realistic nonlocal $NN$
interaction JISP16, seems from Table \ref{bind16} to reproduce well
the binding energies of nuclei with $A\leq 16$.  The lowest state of
natural parity has the correct total angular momentum in each nucleus
studied.  Furthermore, it reproduces the spectra of $^6$Li and
$^{10}$B \cite{JISP16}, which are known to be sensitive to an explicit $NNN$
interaction.  This feature of incorporating by a purely two-body
interaction of what is 
conventionally believed
to be a 3-body force effect, is attributed to the fact that JISP16 is a
nonlocal interaction.  We also note here that JISP16 provides a good
description of the $^6$Li quadrupole moment $Q$ \cite{JISP16} (see
also Table~\ref{t6Li2010} below) that is a recognized challenge due to
a delicate cancellation between deuteron quadrupole moment and the $d$
wave component of the $\alpha$--$d$ relative wave function.
%
%

\section{\boldmath Extrapolation to infinite model space: 
$^{14}$F~ground state}

Recently we introduced an \ {\em ab initio} \ No-Core Full Configuration
(NCFC) approach \cite{NCFC,Bogner:2007rx}, by extrapolating NCSM
results with the bare interaction in successive basis spaces to the
infinite basis space limit.  This makes it possible to obtain basis
space independent predictions for binding energies and to evaluate
their numerical uncertainties.  We use two extrapolation methods: a
global extrapolation based on the results obtained in four successive
basis spaces with five $\hbar\Omega$ values from a 10 MeV interval
(extrapolation A); and extrapolation B based on the results obtained
at various fixed $\hbar\Omega$ values in three successive basis spaces
and defining the most reliable $\hbar\Omega$ value for the
extrapolation.  These extrapolations provide consistent results and were
carefully tested in a number of light nuclei where a
complete convergence can be achieved \cite{NCFC}. 

The NCFC approach was used recently \cite{14F} for the first {\em ab initio}
study of the exotic proton-excess nucleus $^{14}$F. The first experimental
results regarding this four 
proton excess isotope will be available soon from Cyclotron Institute
at Texas A\&M University \cite{Texas}.
The largest calculations were
performed in the $N_{\max}\hbar\Omega$ basis space with $N_{\max} = 8$,
which for this nucleus 
contains 1,990,061,078 basis states with total magnetic projection
$M=0$ and natural parity (negative).  The determination of the lowest
ten to fifteen eigenstates of the sparse Hamiltonian matrix,
for each oscillator parameter $\hbar\Omega$, requires 2 to 3 hours 
on 7,626 quad-core compute nodes at the Jaguar supercomputer at ORNL.

\begin{figure}
\centerline{\includegraphics[width=\columnwidth]{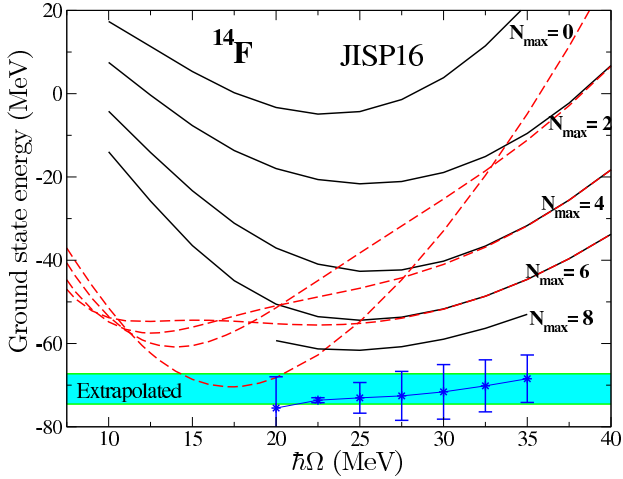}}
\caption{\label{14Ff}  Results for the ground state energy of $^{14}$F
    with bare (solid) and effective (dashed) JISP16 interaction as a
    function of the oscillator parameter $\hbar\Omega$.  The shaded
    area demonstrates the global extrapolation A for the binding
    energy and its uncertainty; the extrapolation B at fixed
    $\hbar\Omega$ is given by stars with 
    its uncertainties indicated by the error bars. The most reliable 
    $\hbar\Omega$ value for this extrapolation method 
    is at $\hbar\Omega=25$ MeV for $^{14}$F.}
\end{figure}

We show our complete results for the $^{14}$F ground state energy in
Fig.~\ref{14Ff}.  The solid curves are our NCSM results with the bare
interaction JISP16.  These results are strict upper bounds for the
ground state energy, and converge monotonically with $N_{\max}$ to the
infinite basis space results.  The dashed curves in Fig.~\ref{14Ff}
are obtained in NCSM calculations with effective $NN$
interactions. 
Again, these results obtained in extremely large NCSM basis spaces and
verified by the new basis-space independent {\em ab initio} NCFC
method, reveal some drawbacks of the effective interaction approach
that was used in the JISP16 fit to binding energies of light nuclei.

By comparing the bare and effective interaction  results in
Fig.~\ref{14Ff}, we observe that the tendency of the LSO(2)
calculations is misleading.  For increasing basis spaces 
from $N_{\max}=0$ to $6$, the minimum of the $\hbar\Omega$-dependent
curves increases, suggesting an approach from below to the infinite
basis space result.  At $N_{\max}=6$, the effective interaction 
produces a nearly flat region at approximately the same energy as the
minimum obtained with the bare JISP16 interaction.  On the other hand,
the bare interaction provides a variational upper bound for the ground
state energy, which decreases with increasing $N_{\max}$.
We see the same convergence pattern in $^{16}$O (see Fig.~\ref{hwdep})
and other $p$ shell nuclei.  For these reasons, we did not perform
expensive $N_{\max}=8$ effective interaction calculations for
$^{14}$F.


We present in Fig.~\ref{14Ff} and in Table~\ref{Egs} the results of
NCFC extrapolations A and B. \ Combining both extrapolation methods
suggests a binding energy of $72\pm 4$ MeV for $^{14}$F. Ironically,
the minimum of the effective interaction calculations in the smallest
$N_{\max}=0$ basis space appears to be closest to the infinite basis
space result.

\begin{table}
  \caption{\label{Egs} 
    NCFC predictions for the ground state energies 
    (in MeV) of $^{13}$O, $^{14}$B and $^{14}$F based on NCSM
 calculations with JISP16 in up to $N_{\max}=8$ basis spaces. We include  in
 parentheses an  estimate of the accuracy of the extrapolations A and
 B. Experimental  data are taken from
 Ref.~\cite{AjzenbergSelove:1991zz}.} 
\begin{center}
    \begin{tabular}{l|ccc}\hline
      Nucleus  &  Extrap. A   &  Extrap. B   &  Experiment \\ \hline
      $^{13}$O & $-75.7(2.2)$ & $-77.6(3.0)$ & $-75.556$ \\
      $^{14}$B & $-84.4(3.2)$ & $-86.6(3.8)$ & $-85.423$ \\
      $^{14}$F & $-70.9(3.6)$ & $-73.1(3.7)$ & --- \\ \hline
    \end{tabular}
\end{center}
\end{table}

To check the accuracy of our approach, we performed similar
calculations for the mirror nucleus $^{14}$B with a known binding
energy of 85.423 MeV \cite{AjzenbergSelove:1991zz}.  This value agrees
with our prediction of $86\pm 4$ MeV.  We also performed NCFC
calculations of the neighboring nucleus $^{13}$O using basis spaces up
to $N_{\max}=8$.  The calculated binding energy of $77\pm 3$ MeV also
agrees with the experimental value of 75.556 MeV
\cite{AjzenbergSelove:1991zz}.

We note that a good description of both $^{14}$F and $^{13}$O in the
same approach is important in order   to have a description of  $^{14}$F
consistent with the experiment in which  $^{14}$F will be produced in
the $^{13}{\rm O}+p$  reaction.  In this way, 
any experimentally observed resonances can be directly compared with
the difference of our predictions for the $^{14}$F and $^{13}$O
energies.  In this respect it is interesting to note that although the
total energies of the extrapolations A and B differ by about 2 MeV,
the differences between the ground state energies of these three
nuclei are almost independent of the extrapolation method: for
$^{14}$F and $^{13}$O the predicted difference is 4.6 MeV, and for
$^{14}$F and $^{14}$B it is 13.5 MeV.  (The numerical uncertainty in
these differences is unclear, but expected to be significantly smaller
than the uncertainty in the total energies.)

\section{Extrapolation to infinite model space: Excitation spectra}

It is also very interesting to calculate the $^{14}$F excitation
spectrum in anticipation of 
the experimental results.  It is unclear how to extrapolate excitation
energies obtained in finite basis spaces, but we can extrapolate the
total energies of excited states using the same methods as
discussed above for the ground state energy.  For the lowest state in
each $J^{\pi}$ channel the convergence pattern should be similar to
that of the ground state; for excited states with the same quantum
numbers we simply assume the same convergence pattern.  We perform
independent separate extrapolation fits for all states.  The
differences between the extrapolated total energies and the ground
state energy is our prediction for the excitation energies.

\begin{table*}
  \caption{\label{t6Li}
    Predictions for the $^{6}$Li ground state $E_{gs}$ and excitation
    energies $E_x$ (in MeV) obtained in different basis spaces with
    JISP16.  For extrapolations A and B we include in parentheses an
    estimate of the accuracy of the  total  energies; 
    for the effective interaction ($V_{\rm eff}$), we present the spread 
    in excitation energy for $\hbar\Omega$ variations 
    from 12.5 to 22.5 MeV.
    Experimental data  (in MeV) are taken from Ref.~\cite{Tilley:2002vg}.}
  \begin{center}    
    \begin{tabular}{l|ccc|ccc|cc}\hline
      & Extrap. A  & Extrap. B  & $V_{\rm eff}$
    & Extrap. A  & Extrap. B  & $V_{\rm eff}$ &\multicolumn{2}{c}{Experiment}\\ 
$E(J^\pi,T)$ 
      &  $N_{\max}=2{-}8$  &  $N_{\max}=4{-}8$  & $N_{\max}=6$ 
      & $N_{\max}=10{-}16$ & $N_{\max}=12{-}16$ & $N_{\max}=14$ 
      & Energy & Width  \\ \hline
$E_{gs}(1^+,0)_1$ 
      & $-30.9(0.6)$   & $-31.1(0.3)$   &       
      & $-31.47(0.09)$ & $-31.48(0.03)$ & 
      & $-31.994$ & Stable\\ \hline
$E_{x}(3^+,0)$ 
      & 2.6(0.5) & 2.5(1.2) & 2.2--2.7 
      & 2.56(0.04) & 2.55(0.07) & 2.53--2.55
      & 2.186 & $24\cdot 10^{-3}$\\
$E_{x}(0^+,1)$ 
      & 3.6(0.6) & 3.5(1.2) & 3.3--3.7  
      & 3.68(0.06) & 3.65(0.06) & 
                                  3.6--3.8 
                                         & 3.563 &$8.2\cdot 10^{-6}$\\
$E_{x}(2^+,0)$ 
      & 5.3(0.9) & 5.5(1.8) & 4.8--5.8  
      & 4.5(0.1) & 4.5(0.2) & 4.8-5.0 
                                          & 4.312 & 1.30\\
$E_{x}(2^+,1)$ 
      & 6.3(0.7) & 6.1(1.6) & 6.2--6.5  
      & 5.9(0.1) & 5.9(0.1) & 6.0--6.4 
                                          & 5.366 & 0.54\\
$E_{x}(1^+,0)_2$ 
      & 6.1(1.7) & 6.6(0.3) & 7.1--8.5  
      & 5.4(0.3) & 5.4(0.2) & 6.1--6.6
                                          & 5.65 & 1.5 \\ \hline
    \end{tabular}
  \end{center}
\end{table*}

\begin{table*}
  \caption{\label{t14F-B}
  Predictions for the $^{14}$F and $^{14}$B excitation energies $E_x$
 (in MeV) based on NCSM calculations with JISP16 in up to $N_{\max}=8$
 basis spaces.  See Table \ref{t6Li} for details.
Experimental data  (in MeV) are taken from Ref.~\cite{AjzenbergSelove:1991zz}.}
  \begin{center}
    \begin{tabular}{l|ccc|ccc|cc}\hline
    &  \multicolumn{6}{c|}{
 {\it Ab initio} NCFC and effective interaction NCSM  calculations with JISP16}
&\multicolumn{2}{c}{Experiment}\\
    &  \multicolumn{3}{c|}{$^{14}$F}&\multicolumn{3}{c|}{$^{14}$B} 
                                   & \multicolumn{2}{c}{$^{14}$B}\\
$E(J^\pi,T)$  & Extrap. A & Extrap. B & $V_{\rm eff},,\:N_{\max}=6$
  &  Extrap. A & Extrap. B & $V_{\rm eff},\:N_{\max}=6$
            &$J^\pi$ &Energy  \\ \hline
$E_x(1^-,2)_1$  &0.9(3.9) & 1.3(2.5)  & 1.4--2.2 
        & 1.1(3.5) & 1.4(2.8) &  1.4--2.3 
                                                   & $(1^-)$ &0.74(4)  \\
$E_x(3^-,2)_1$ & 1.9(3.3) & 1.5(4.6) & 1.0--1.8  
             &1.7(2.9) & 1.4(4.6) & 1.0--2.1 
                                                    & $(3^-)$ & 1.38(3)  \\
$E_x(2^-,2)_2$ & 3.2(3.5) & 3.3(3.5) &  3.3--3.7  
    &3.3(3.1) & 3.3(3.8) & 3.5--3.8 
                                                    & $2^-$ & 1.86(7) \\
$E_x(4^-,2)_1$ &3.2(3.2) & 2.8(4.8) & 2.0--2.6   
      & 3.1(2.9) & 2.7(4.8) & 2.0--3.1 
                                                   & $(4^-)$ & 2.08(5)\\
& & & & & & & ? &2.32(4)\\
& & & & & & & ? &2.97(4)\\
$E_x(1^-,2)_2$ & 5.9(3.5) & 5.4(4.6) & 5.8--6.4 
               & 5.9(3.1) & 5.5(4.8) & 5.7--6.4  
                                                    \\
$E_x(0^-,2) $  & 5.1(5.4) & 5.8(1.0) & 5.8--10.5 
               & 5.5(4.8) & 6.1(1.4) & 4.9--10.4 
                                                   \\
$E_x(1^-,2)_3$ & 6.2(4.8) & 6.3(2.8) & 7.2--11.5 
               & 6.4(4.3) & 6.4(3.1) & 6.1--11.3  
                                                    \\
$E_x(2^-,2)_3$ & 6.4(4.6) & 6.3(3.4) & 7.3--10.9  
               & 6.9(4.1) & 6.7(3.6) & 6.6--10.9  
                                                    \\ 
$E_x(3^-,2)_2$ & 6.9(4.2) & 6.4(4.6) & 7.6--10.6 
               & 7.0(3.7) & 6.5(4.7) & 6.4--10.5 
                                                    \\
$E_x(5^-,2)$ & 8.9(3.5) & 7.9(5.9) &  9.2--11.0 
               & 8.8(3.1) & 7.8(5.9) & 8.5--10.8 
           \\ \hline
    \end{tabular}
  \end{center}
\end{table*}

To verify this approach to calculate the NCFC excitation energies, we
apply it to the spectrum of $^6$Li nucleus, see Table~\ref{t6Li}. 
We have results for
$^6$Li in basis spaces up to $N_{\max} =16$ where a good convergence
is achieved and hence the extrapolation uncertainties are small.
These results are compared in Table~\ref{t6Li} with the extrapolations
based on calculations in basis spaces up to $N_{\max} = 8$, i.~e. in
the same basis spaces used for the $^{14}$F and $^{14}$B studies.

We see that the excitation energies based on $N_{\max} = 8$ and
smaller basis space results are consistent with the results obtained
in larger spaces.  The level ordering is the same \ and \ the \
difference \
between \ the \ $N_{\max} = 8$ \ and ${N_{\max} = 16}$ results is generally
much smaller than the estimated uncertainties in the total energies
of the ${N_{\max}= 8}$ extrapolations.  This suggests that the numerical
uncertainty in the excitation energies is significantly smaller than
the uncertainty in the total energies: apparently, the calculated
total energies share a significant systematic uncertainty, an
overall binding  uncertainty, which
cancels when results are expressed as excitation energies.
Furthermore, we see that both extrapolation methods agree very well
with each other (within their error estimates), and that the error
estimates decrease as one increases the basis space.

The two lowest excited states in $^6$Li are narrow resonances. Our
predictions for these states obtained by extrapolations A and B and
with effective interaction, agree very well with experiment. The bare
and effective interaction excitation energies of these states 
show very little
dependence on $\hbar\Omega$. 

On the other hand, the three higher excited states have a much larger
width, see Table~\ref{t6Li}.  Our calculations for these broad
resonances show a significant dependence on both $\hbar\Omega$ and
$N_{\max}$, in particular for the excited $(1^+,0)_2$ state which has
the largest width.  The extrapolation B to infinite model space
reduces but does not eliminate completely the $\hbar\Omega$ dependence.  We
further note that the $\hbar\Omega$-dependence of these excitation
energies is typical for wide resonances as observed in comparisons of
NCSM results with inverse scattering analysis of $\alpha$-nucleon
scattering states~\cite{inverse}, and that the slope of the
$\hbar\Omega$ dependence increases with the width of the resonance.
Thus, there appears to be a significant correlation between
the resonance width and the $\hbar\Omega$ dependence.  The validity of
the extrapolation to infinite model space is not entirely clear for
these states.

We noted earlier that the effective  interaction does not
provide a monotonic approach to the infinite basis space for the
binding energies and this prevents simple extrapolation.  On the other
hand, the excitation energies with the effective  interaction 
are often quite stable with $N_{\max}$.  However, it is important to
realize that this does not necessarily mean that these excitation
energies are numerically converged: they can be $\hbar\Omega$-dependent.
The dependence of the excitation energies on $\hbar\Omega$ decreases
slowly with increasing $N_{\max}$, as seen in
Table~\ref{t6Li}.  In fact, the excitation energies obtained with
effective  interaction based on JISP16 are nearly the same as those
obtained with the bare JISP16 interaction.   For most states, the NCFC 
provides better results for the excitation
energies, with less basis space dependence than the effective  interaction NCSM
calculations in finite basis spaces.  Nevertheless, we can employ the
effective  interaction to obtain estimates of the binding and excitation
energies when only  small basis spaces are attainable and the NCFC
extrapolations are impossible.

We summarize our results for the spectra of $^{14}$F and $^{14}$B in
Table~\ref{t14F-B}.  The excitation energies are obtained as a
difference between the extrapolated total energies of the excited
state and that of the ground state (see Table~\ref{Egs}).
The spectra are rather dense and the spacing between
energy levels is smaller than the quoted numerical uncertainty, which
is that of the extrapolated total energies of the excited states.
However, as discussed above, we expect that for narrow resonances the
actual numerical error in the excitation energy is (significantly)
smaller than the error in the total energy.

\begin{figure}
\centerline{\includegraphics[width=\columnwidth]{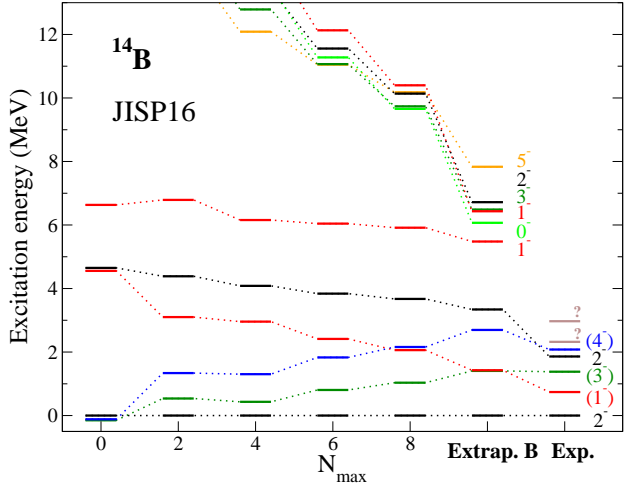}}
\caption{\label{14Bspectr} Negative parity $^{14}$B spectrum obtained with
  bare  JISP16 interaction at fixed $\hbar\Omega=25$~MeV in successive
 basis spaces,  and extrapolated to infinite basis space using Extrapolation B.
    Experimental data are taken from Ref.~\cite{AjzenbergSelove:1991zz}.}
\end{figure}

Figure~\ref{14Bspectr} shows that different excited states can have
very different convergence behavior. We present in
Fig.~\ref{14Bspectr} the $^{14}$B results and we note that the behavior of the
$^{14}$F states is similar. There are five low-lying excited states with \
excitation \ energies \ showing \ a \ weak \ dependence,\  within about a 1.5 MeV range, 
on the basis space as $N_{\max}$
increases from 
$2$ to $8$.  Then there are numerous higher excited states which
depend strongly on the basis space: their excitation energies decrease
rapidly with increasing $N_{\max}$.  Only after extrapolation to the
infinite basis space do they appear at excitation energies comparable
to the other low-lying excited states.  
The lowest five excited states have a weak dependence on $\hbar\Omega$, 
whereas the higher excited states depend
strongly on it and this strong $\hbar\Omega$ dependence is manifested in
larger global 
extrapolation A  uncertainties obtained as a spread of 
extrapolated results at different $\hbar\Omega$ values.  We expect our results for
these higher 
excited states to have a larger numerical error than our results for
the lower excited states with the weaker $\hbar\Omega$ dependence.
Furthermore, in analogy to the excited states in $^{6}$Li
discussed above, we expect these higher states to be broad resonances.
Interestingly, the high-lying $J^{\pi}=5^-$ state has a relatively
weak $\hbar\Omega$ dependence (compared to states with similar
excitation energy); it  is also less dependent on $N_{\max}$, and may
correspond to a narrower resonance.

The excitation energies of the lowest five and the  high-lying
$J^{\pi}=5^-$ state obtained with effective interaction are less
$\hbar\Omega$-dependent than the other states shown and are consistent
with the NCFC extrapolations. For 
the higher excited states, the NCFC results differ significantly from the 
 effective interaction predictions;  these extrapolated results also
 tend to have a somewhat 
weaker dependence on $\hbar\Omega$ than the results in finite basis
spaces, and are  expected to be more accurate.

Some of the excited states in $^{14}$B were observed experimentally.
Unfortunately, the spin of most of these states is doubtful or unknown.
Overall, our predicted excitation energies appear to be too large when compared
with the experimental data; in particular our prediction for the
excited $2^-$ state, the only excited state with a firm spin
assignment, is about 1.5 MeV above the experimental value.  However,
the spin of the lowest five states agrees with experiment, except for
the $2^-$ and $4^-$ being interchanged, assuming that the tentative
experimental spin assignments are correct.  We do not see additional
states between 2 and 3 MeV, but this could be related to the fact that
all our excitation energies appear to be too large. 
It would also be very interesting to
compare our predictions for the $^{14}$F binding energy and spectrum
with the experimental data that are anticipated soon.

\section{Towards a renewed JISP16}

As we have seen above, the effective interaction approach used in the
fitting of the JISP16 interaction, can be misleading in evaluating
binding energies of nuclei. The new {\em ab initio} NCFC aproach
provides much more reliable predictions for bindings. Therefore the
results presented in Table \ref{bind16} should be reevaluated using
extrapolations A and~B. The NCFC binding energies of some nuclei
obtained with JISP16  are  presented  in  Table 
\ref{tabA3-16}. \ The  NCFC 
approach clearly reveals a deficiency of the JISP16 interaction: it
overbinds essentially nuclei with mass $A\geq 14$ 
and $N \sim Z$.

\begin{table*}
\caption{Binding energies (in MeV) of some nuclei obtained with JISP16 and
JISP16$_{2010}$ $NN$ interactions by extrapolations A and B; the~$N_{\max}$
columns show the largest model space used for the extrapolations.}
\label{tabA3-16}
\begin{center}
\begin{tabular}{c|c|ccc|ccc}\hline
 & &\multicolumn{3}{|c|}{JISP16} &\multicolumn{3}{|c}{JISP16$_{2010}$}
\\\cline{3-8}
Nucleus      & Experim.  & Extrap. A &  Extrap. B  & $N_{\max}$
          & Extrap. A &  Extrap. B  & $N_{\max}$\\ \hline
$^{3}$H   &  8.482 &  $8.369\pm 0.001$ &  $8.3695\pm 0.0025$  & 18
  & $8.369\pm 0.010  $& $8.367_{-0.007}^{+0.012}$ & 14 \\ 
$^{3}$He  &  7.718 &  $7.665\pm 0.001$ &  $7.668\pm 0.005$  & 18  
 & $7.664\pm 0.011$ &$7.663\pm 0.008$ & 14\\
$^{4}$He       & 28.296 & $28.299\pm 0.001$ &  $28.299\pm 0.001$ & 18 
 &$28.294\pm 0.002$ &$28.294_{-0.001}^{+0.002}$ & 14 \\
$^8$He & 31.408 & $29.69\pm 0.69$ & $29.29\pm 0.96$ &10
     & $30.30\pm 0.46$ & $29.99^{+1.31}_{-1.06}$ & 10 \\ 
$^{6}$Li
          & 31.995 & $31.47\pm 0.09$ & $31.48\pm 0.03$  &16 
 &$31.33\pm 0.16$ & $31.34\pm 0.07$ &14 \\
$^{10}$B        &64.751 &$63.1\pm 1.2$ &$63.7\pm 1.1$ &8
 &$62.6\pm 1.4$ &$63.4\pm 1.5$ &8 \\
$^{12}$C        & 92.162 & $93.9\pm 1.1$ & $95.1\pm 2.7$  & 8
 &$91.1\pm 1.3$ & $92.3\pm 2.9$ & 8\\
$^{14}$C        & 105.284 & $112.1\pm 2.1$ &$114.3\pm 6.0$ & 8
 &$102.5\pm 1.6$ & $104.8\pm 3.6$ & 8 \\
$^{14}$N        & 104.659 & $114.2\pm 1.9$ & $115.8\pm 5.5$ & 8
 &$102.7\pm 1.5$ & $104.7\pm  3.1$ & 8 \\
$^{16}$O      &127.619 & $143.5\pm 1.0$ & $150\pm 14$ & 8 
 &$126.7\pm 3.1$ &$129.6\pm 6.1$ &8  \\
\hline
\end{tabular}
\end{center}
\end{table*}

\begin{table*}
\caption{PET rotation parameters $\vartheta$ in various $NN$ partial waves
  defining the JISP16$_{2010}$  $NN$ interaction.}  
\label{JISP-par}
\begin{center}
\begin{tabular}{c|ccccccccc}\hline
Partial wave &$^1s_0$              &$^3p_0$       &$^1p_1$        &$^3p_1$ 
&$^3s_1{-}{^3d}_1$ &$^1d_2$      &$^3d_2$ &$^3p_2{-}{^3f}_2$ &$^3d_3{-}{^3g}_3$
 \\ \hline
$\vartheta$ & $-0.0966^\circ$ &$-8.72^\circ$ &$-15.62^\circ$ &$-6.01^\circ$ 
   &$-11.00^\circ$ &$-2.73^\circ$ &$7.25^\circ$ &$7.00^\circ$  &$0.457^\circ$
\\ \hline
\end{tabular}
\end{center}
\end{table*}

These deficiencies of the $NN$ interaction can be addressed by a new fit of
the PET parameters defining JISP interaction based on the NCFC
calculations. Such a fit is much more complicated since it requires
calculations in a number of successive basis spaces for
each nucleus and each  set of parameters. The renewed 
$NN$ interaction obtained in this fit which 
we refer to as  JISP16$_{2010}$,  is fixed by the set of PET rotation
parameters listed in Table~\ref{JISP-par} (the definition of PET rotation
parameters $\vartheta$ can be found in Refs.~\cite{ISTP,JISP6}). The
JISP16 and JISP16$_{2010}$  interactions are characterized by 
the same $\vartheta$ value in coupled $sd$
partial waves; hence both interactions predict the same deuteron
properties (see Table~\ref{d-prop}). All the remaining $\vartheta$ values
listed in Table~\ref{JISP-par} differ between 
JISP16$_{2010}$ and JISP16. We note also that the JISP16 interaction was
defined only in the $NN$ partial waves with total momenta $J\leq 4$
while the JISP16$_{2010}$  interaction is extended to all partial waves
with  $J\leq 8$. The description of $NN$ scattering data by
JISP16$_{2010}$ and JISP16 interactions is the same since they are
related by phase-equivalent transformations. 

We compare  binding energies obtained with JISP16 and JISP16$_{2010}$
interactions in Table~\ref{tabA3-16}. It is seen that the new
interaction essentially improves the description of the $p$ shell nuclei. In
particular, JISP16$_{2010}$ provides nearly exact binding energies of
nuclei with $10\leq A\leq16$ and only slightly underbinds some of lighter
nuclei listed in Table~\ref{tabA3-16}.

\begin{table*}
\caption{NCFC predictions of the $^{6}$Li ground state $E_{gs}$ and
 excitation  $E_x$ energies (in MeV) in comparison with Green's function
 Monte Carlo (GFMC) results obtained with Argonne AV18 $NN$ and Urbana
 UIX and Illinois IL2 $NNN$ interactions. 
We present in 
 parentheses the estimate of the accuracy of extrapolations A and B of the
 absolute energies of excited states. For the ground state rms point-proton
 radius $r_p$ and quadrupole moment $Q$, we present the interval of
 values (in fm and $e\cdot \rm fm^2$, respectively) 
 obtained in the largest model spaces  with bare interactions 
 with $\hbar\Omega=15{-}25$~MeV, i.~e. in the same interval of $\hbar\Omega$
 values that was used in extrapolation~A for the ground state. 
}
\label{t6Li2010}
\begin{center}

\begin{tabular}{l|cc|cc|cc|c}\hline
& \multicolumn{2}{c|}{JISP16} & \multicolumn{2}{c|}{JISP16$_{2010}$} 
                                            &AV18+UIX &AV18+IL2  \\
  & Extrap. A  & Extrap. B & Extrap. A & Extrap. B 
               &  GFMC \cite{GFMC,IL2exc} &GFMC \cite{Illinois,IL2exc} &\\
$E(J^\pi,T)$ &   $N_{\max}=10{-}16$ &  $N_{\max}=12{-}16$  
      & $N_{\max}=8{-}14$ &$N_{\max}=10{-}14$ & & & Experim.  \\ 
 \hline
$E_{gs}(1^+,0)_1$ & $-31.47(0.09)$ & $-31.48(0.03)$ 
        & -31.33(0.16) & -31.34(0.07) &$-31.25(8)$  &$-32.0(1)$&$-31.994$ \\
$r_p$   &\multicolumn{2}{c|}{2.137--2.240 
}
        &\multicolumn{2}{c|}{2.109--2.225} &2.46(2) &2.39(1) & 2.32(3) \\
$Q$   &\multicolumn{2}{c|}{$-0.071{-}\,{-0.075}$}
        &\multicolumn{2}{c|}{$-0.081{-}\,{-0.102}$}
                               &$-0.33(18)$ &$-0.32(6)$& $-0.082(2)$ \\ \hline
$E_{x}(3^+,0)$ & 2.56(0.04) & 2.55(0.07) 
        &2.08(0.16) &2.097(0.003)  & 2.8(1) &2.2 & 2.186\\
$E_{x}(0^+,1)$ & 3.68(0.06) & 3.65(0.06) 
        &3.46(0.18) &3.498(0.007) &3.94(23) &3.4 & 3.563 \\
$E_{x}(2^+,0)$ & 4.5(0.1) & 4.5(0.2) 
        & 4.5(0.3) &4.39(0.16) & 4.0(1) &4.2 & 4.312 \\
$E_{x}(2^+,1)$ & 5.9(0.1) & 5.9(0.1) 
        & 5.8(0.3)&5.72(0.05)  & &5.5 & 5.366 \\
$E_{x}(1^+,0)_2$ & 5.4(0.3) & 5.4(0.2) 
        &5.5(0.7) & 5.72(0.25) & 5.1(1) &5.6  & 5.65\\ \hline
\end{tabular}
\end{center}
\end{table*}

Table~\ref{t6Li2010}  presents the results of $^6$Li properties
calculations 
with JISP16 and JISP16$_{2010}$. It is seen that JISP16 and
JISP16$_{2010}$  provide
more or less the same quality of $^6$Li properties:
the binding energy is better described by JISP16  but  the 
excitation spectrum is  are 
better  described  by JISP16$_{2010}$. Both JISP16 and JISP16$_{2010}$
describe $^6$Li 
with a quality similar to the  Argonne AV18 $NN$ interaction in combination
with Urbana UIX $NNN$ force.
 Both JISP16 and JISP16$_{2010}$ also provide a comparable description of
 $^6$Li as  AV18 
in combination with Illinois IL2 $NNN$ interaction: AV18+IL2
is more accurate for the binding energy prediction, overestimates the
rms point-proton radius by approximately the same value as the
underestimation of $r_p$ by JISP16 and JISP16$_{2010}$  interactions (note
however that the  $r_p$ value is not completely converged in our
calculations and still increases with $N_{\max}$), provides
more or less the same accuracy for excitation energies, but the
 $^6$Li quadrupole moment calculated with AV18+IL2 is
significantly larger than the experiment.

We plan to explore the properties of the refined realistic
nonlocal $NN$ interaction JISP16$_{2010}$  in systematic large-scale
calculations of other light nuclei including the ones with $A>16$ and
away from $N\sim Z$,
and to carefully study its predictions not 
only for the binding energies but also for the spectra, electromagnetic
transitions and other observables. Our plan is also to tune the
interaction to the description of 
nuclear matter properties.

\section{Concluding remarks}

We believe 
that JISP16$_{2010}$ has  good prospects to success in the nuclear
structure studies. This $NN$ interaction provides a  high-quality
description of the $NN$ data together with a very 
reasonable description of many-body nuclear systems without referring to
$NNN$ forces. Moreover, a very specific form of this interaction~---
small-rank matrix in the oscillator basis with a reasonable
$\hbar\Omega$ value~--- is responsible for a fast convergence of shell
model calculations which makes it possible to rely on bare interaction
results and extrapolate them to infinite model space. 
We consider this  $NN$ interaction  as an important realistic
ingredient of~the~new accurate {\em ab initio} NCFC approach in nuclear
structure theory. 

In constructing JISP-type $NN$ interaction models, we adopted 
only the accepted symmetries of the strong interaction and neglected 
explicit constraints such as the long-range behavior from 
meson-exchange theory. However, this does
not mean that the JISP16 and JISP16$_{2010}$  interactions are inconsistent with
meson-theoretical forms of the $NN$ interaction. On the contrary, it is 
well-known that the one-pion exchange  dominates the $NN$ 
interaction in higher partial waves 
and the long-range behavior of $NN$ interaction
in lower partial waves. In this context, we showed in Ref. \cite{ISTP}
that our scattering wave functions in higher partial waves are nearly
indistinguishable from those of the Nijmegen-II meson-exchange 
potential. Also, in lower partial waves, our wave functions are very close
to those of  Nijmegen-II at large distances and a small difference 
is seen only at higher energies. 
Finally, we introduced the PETs of JISP16 and JISP16$_{2010}$   only  
in lower partial waves and only in a 
few lowest oscillator components of the potential 
with a large value of $\hbar\Omega=40$ MeV.
As a result, PETs reshape the wave functions at
short distances ($\lesssim 1$~fm) only.  Thus, the JISP-type interactions
appear to be consistent with the well-established pion-exchange tail as
embodied in the Nijmegen-II $NN$ interaction. 

We propose our JISP16$_{2010}$  as a realistic  $NN$ interaction
describing the two-body observables with 
high precision and providing a reasonable economic 
description of properties of many-body nuclear systems 
in  microscopic {\em ab initio} approaches. The economy arises from
the softness of the interaction represented in a separable oscillator form.
Short distance phase-equivalent transformations adjust
the off-shell properties successfully to reduce the roles of multi-nucleon
interactions. The particular mechanism of this reduction is still not
completely clear. However, our results clearly demonstrate that such a
mechanism exists and should be studied in detail.

\section*{Acknowledgments}

Considerable  development \ of  parallel  algorithms \ in  the 
code MFDn \cite{MFDn,MFDn-WEB} \
enabled the calculations presented here.  The most recent developments 
\cite{ACM,SSNV,SciDAC09,Laghave}
enabled the largest calculations to be performed on leadership
class computers. 

We thank Vladilen Goldberg (Texas A\&M University) for very valuable
discussions.  We also thank Esmond Ng, Chao Yang and Philip Sternberg
of LBNL and Masha So\-son\-kina of Ames Laboratory for fruitful
discussions on computational science and applied mathematics issues
underlying code developments.  This work was supported by the US DOE
Grants DE-FC02-09ER41582 and DE-FG02-87ER40371 and the FAO Contract
P521.  Computational resources were provided by DOE \ through \ the
National Energy \ Research \ Supercomputer \ Center \ (NERSC) \ and through an
INCITE award (David Dean, PI)

\end{document}